\documentclass[a4paper]{jpconf}
\usepackage{graphicx}

\usepackage{array}

\usepackage{url}  
\usepackage{hyperref} 
\begin{document}
\title{Using Jupyter Notebooks to foster computational skills and professional practice in an introductory physics lab course 
}

\author{E. Tufino$^{1}$, S. Oss$^{1}$, and M. Alemani$^2$}

\address{$^1$ Department of Physics, University of Trento, 38123 Trento, Italy}
\address{$^2$ Physics and Astronomy Institute, University of Potsdam, Potsdam 14476, Germany}

\ead{eugenio.tufino@unitn.it, stefano.oss@unitn.it, alemani@uni-potsdam.de}

\begin{abstract}
In this paper, we detail the integration of Python data analysis into a first-year physics laboratory course, a task accomplished without significant alterations to the existing course structure. We introduced tailored laboratory computational learning goals and designed activities to address them. We emphasise the development and application of Jupyter Notebooks, tailored with exercises and physics application examples, to facilitate students' mastery of data analysis programming within the laboratory setting. These Notebooks serve as a crucial tool in guiding students through the core principles of data handling and analysis in Python, while working on  simple experimental tasks. The results of the evaluation of this intervention offer insights into the advantages and challenges associated with early integration of computational skills in laboratory courses, providing valuable information for educators in the field of physics education. This study demonstrates a practical and effective way of embedding computational skills into the physics curriculum, and contributes to the ongoing efforts of the physics education research community.

\end{abstract}

\section{Introduction}\label{Introduction}
In the context of today's technological advancement, computational skills in physics are essential. The increased data generation in experimental sciences necessitates robust data analysis tools. This urgency is echoed in the physics education community’s efforts to integrate computation into curricula and by recommendations from the American Association of Physics Teachers (AAPT) \cite{Behringer} and the Partnership for Integration of Computation into Undergraduate Physics (PICUP) \cite{PICUP}. 

Following these recommendations a variety of approaches to incorporating computation have been developed, ranging from adapting existing physics lectures to include computational elements \cite{Chabay, caballero2014, odden2021}, to the creation of project-based computational courses \cite{Burke} and the integration of computational tools into laboratory settings. For instance, in introductory first-year laboratory courses, tools like Microsoft Excel \cite{Sachmpazidi} and VPython \cite{Serbanescu} modules have been utilised, while more advanced settings often employ programming languages such as Python \cite{Serbanescu}. For an extensive and current (2023) overview of the role of computation in physics education, refer to the Resource Letter mentioned in \cite{Atherton}.


This paper examines the introduction of Python for data analysis in a first-semester laboratory course. Python was chosen for its widespread use, market demand, flexibility, and comprehensive online documentation. Our goal aims to make programming more accessible to all physics major students from the outset of their university education, lowering the entry barrier to programming and fostering active student engagement. The challenges of embedding computational tools into already intensive first-semester laboratory courses, which focus on statistical analysis, measurement uncertainties, and experimental documentation, are addressed.

In this conference proceeding, the teaching approach, the definition of computational learning goals and the curriculum structure are presented and discussed. We showcase examples of developed Jupyter Notebooks (JNs) \cite{jupyter}, illustrating the pedagogy implemented. These examples highlight the integration of computational tools into the curriculum and the methods used to engage students. While we explain the teaching approach and curriculum development in this paper, a comprehensive analysis of the assessment outcomes is available in an article submitted to a scientific journal, offering an in-depth evaluation of the effectiveness of these methods in fostering computational skills. A preliminary version of this article is available on arXiv \cite{arXivPython}.

\section{Our approach for integrating computational skill in the physics laboratory course
}\label{Design Approach}
The design and implementation of our intervention involved three main steps. Initially, we defined specific laboratory computational learning goals, focusing on introductory Python data analysis skills. We then developed curricular materials to support these goals and facilitate student learning. The final step involved evaluating the achievement of these goals through analysis of student work and pre-post questionnaires.

We implemented this intervention in the University of Potsdam's first-semester laboratory course for physics majors. This course is part of a progressive series of four modules during the first four semesters. The complete sequence was redesigned from 2016 to focus on developing experimental skills and encouraging expert-like thinking. Details of the course redesign and its assessment are presented elsewhere \cite{Alemani_course, Alemani_GECLASS}. Note that while introducing computational aspects into the course, we retained the original laboratory goals, slightly reducing in-class exercises and replacing the use of a scientific data analysis software (SciDavis) with Python programming for data analysis \cite{scidavis}. 

The laboratory computational learning goals for Python were informed by the AAPT recommendations and relevant literature on computational thinking \cite{Behringer,caballero2014, odden2021,weller2022}, focusing on practical data analysis skills. 
Our goals were categorised into two areas, that we refer to as  'Implement' and  'Communicate'. 
In the first, we aim to enable students to write Python codes for problem-solving and for manipulating, analysing, and visualising data sets. In the second, we want students to learn how to effectively communicate their work using Jupyter Notebooks, write clear and well-commented Python code and create clear, informative, and comprehensive graphs using Python.

Our curriculum design for integrating computation involved using JNs for Python programming, which were applied by students also in experimental activities and tested in subsequent semesters.  We created introductory and advanced JNs for both collaborative and independent learning. The material, accessible on GitHub \cite{etufino_github}, was tailored to be approachable even for students without prior coding experience.

The assessment of the achievement of the defined computational learning goals involved both formative and summative methods, and students' attitudes towards computation were surveyed. Importantly, this evaluation was distinct from the assessment of the existing laboratory learning goals, which were evaluated separately as in the previous years.

We considered emotional and attitudinal factors, such as students' confidence and self-efficacy \cite{bandura1986}, as well as their sense of authenticity in using Python as practising physicists do. Our primary aim was to lower the barrier to programming, cultivating a belief among all students, regardless of their prior experience, that they could actively engage in Python data analysis. Furthermore, we encouraged an authentic attitude of reading and understanding the codes written by others and building upon them, aligning with the practices of professional physicists. Our assessment encompassed surveys and ongoing monitoring during laboratory work. For instance, students were encouraged to perform data analysis directly while doing experiments in the laboratory sessions. Additionally, we dedicated effort to nurturing a positive attitude and self confidence among students in their use of Python for data analysis, ultimately enhancing their computational skills.

\begin{table}[]
\caption{\label{Table1}Overview of the in class activities and data collection tool used to introduce Python data analysis in the course and to assess students skills acquisition.} 

\begin{center}

\renewcommand{\arraystretch}{1.8}

\begin{tabular}{ m{2.2cm}  m{3cm} m{5cm}  m{2.5cm} } 
\br

Timeline& \raggedright Data Collection Tool & Description of the Tool & Dimension Evaluated \\
\mr

Beginning of first semester &Pre-Survey & Pre-course survey completed individually by students. & Attitudinal and Emotional\\
\raggedright Session 1 and \newline Session 2 & \raggedright Jupyter Notebooks & \raggedright Students work in groups, completing and submitting JNs (1-4) with their solutions to exercises.&Computational learning goals\\
\raggedright First and second laboratory activities & \raggedright Laboratory notebook & Students in groups submit either a JN or a lab notebook containing the data analysis of the activity. & Computational learning goals and lab learning goals\\ \raggedright After laboratory activities & Post-Survey & Post-course survey completed individually by students. & Attitudinal and Emotional\\
\raggedright Beginning of second semester exercises & \raggedright Jupyter Notebooks & \raggedright Students in groups submit either a JN or a lab notebook containing the data analysis of the activity. & Computational learning goals and lab learning goals\\

\br
\end{tabular}
\end{center}
\end{table}

Table \ref{Table1} provides an overview of the activities and assessment tools employed in this study. It includes a description of each tool and specifies the dimension being evaluated. The data were collected from 48 students who voluntarily participated in the study. The pre- and post-course surveys were completed individually, with 42 matched responses obtained for analysis. Other tasks, such as Jupyter Notebook activities and lab notebook submissions, were conducted in group settings throughout the course.

\section{Description of the Jupyter Notebooks developed
}\label{Examples JNs}

In this section, we provide a detailed description of the JNs that were prepared by us for first year introductory laboratory course students. Our approach is to teach  programming concepts within the context of physics, using examples related to calculating physical quantities or simulating simple physical systems, thereby enabling students to directly relate programming to their field of study. Students learn to import, manipulate, and analyse data using Python. They implement using Python techniques like linear regression, error analysis, and graphical data representation, previously introduced and reinforced throughout the course. Students  practise these concepts and techniques in real-world scenarios.
The curriculum is divided into three modules, each focusing on a specific aspect of Python data analysis: 

\subsection{Introduction to Python (four JNs):}
This module provides a comprehensive introduction to Python programming, covering essential concepts such as installation, basic calculations, text formatting, LaTeX usage, and creating simple plots. Furthermore, the implementation of the least squares method for linear regression and the calculation of statistics parameters like the coefficient of correlation and the chi-squared are treated.  After this introduction, done in two (3h each) in class sessions, students engage in two in-person lab sessions, each lasting three hours, to gain hands-on experience with these data analysis concepts and techniques. 

\subsection{Application Examples (five JNs):}
This module applies Python data analysis to real-world physics phenomena. Students explore exercises that involve calculating the moving average applied to sunspot time-series, plotting and analysing data of light source illuminance vs. distance, creating histograms of the period of oscillation of a pendulum, performing a simulation of the parabolic motion, and plotting atmospheric $CO_2$ data from a website. These exercises are completed asynchronously and independently to allow students to explore Python's capabilities in a self-paced manner. 

\subsection{Additional Python Techniques (two JNs):}
This module focuses on mastering data import and manipulation techniques across various platforms. Students learn how to effectively import and manipulate data files using Jupyter Lab-Anaconda, Google Colab, and online sources. These techniques are acquired through asynchronous independent work, allowing students to further extend their abilities and apply them to future projects. 

\vspace{0.3cm}
Through this structured curriculum, students can gain a solid foundation in Python data analysis, enabling them to effectively analyse and interpret real-world data in their physics studies and beyond. Note that our intervention did not significantly alter the course structure.
The following Section  describes sample activities with Jupyter Notebooks used in the course.

\section{Examples of Jupyter Notebooks}
\subsection{Lesson 2's second JN: exploring Pandas and the least squares method}
In the second lesson dedicated to Python (Notebook 4), we introduced the use of  Pandas, a Python library widely used in professional data analysis. This session built on the students' existing knowledge of graphing data and of the least squares method, marking a transition from traditional paper and pencil method to a computational approach. The notebook starts with a comprehensive introduction to Pandas, including practical guidance on how to import data into Pandas DataFrames and highlighting its crucial role in physics data analysis. A central part of the notebook is an exercise focused on the motion of a vertically thrown stone (see figure \ref{fig2}). In this exercise, students are asked to analyse the stone's trajectory, aiming to test the applicability of the relationship between the stone velocity $v$ and its height $h$ as $v^2=2gh$, where g is the acceleration due to gravity.

\begin{figure}
\begin{center}
\includegraphics[width=6in]{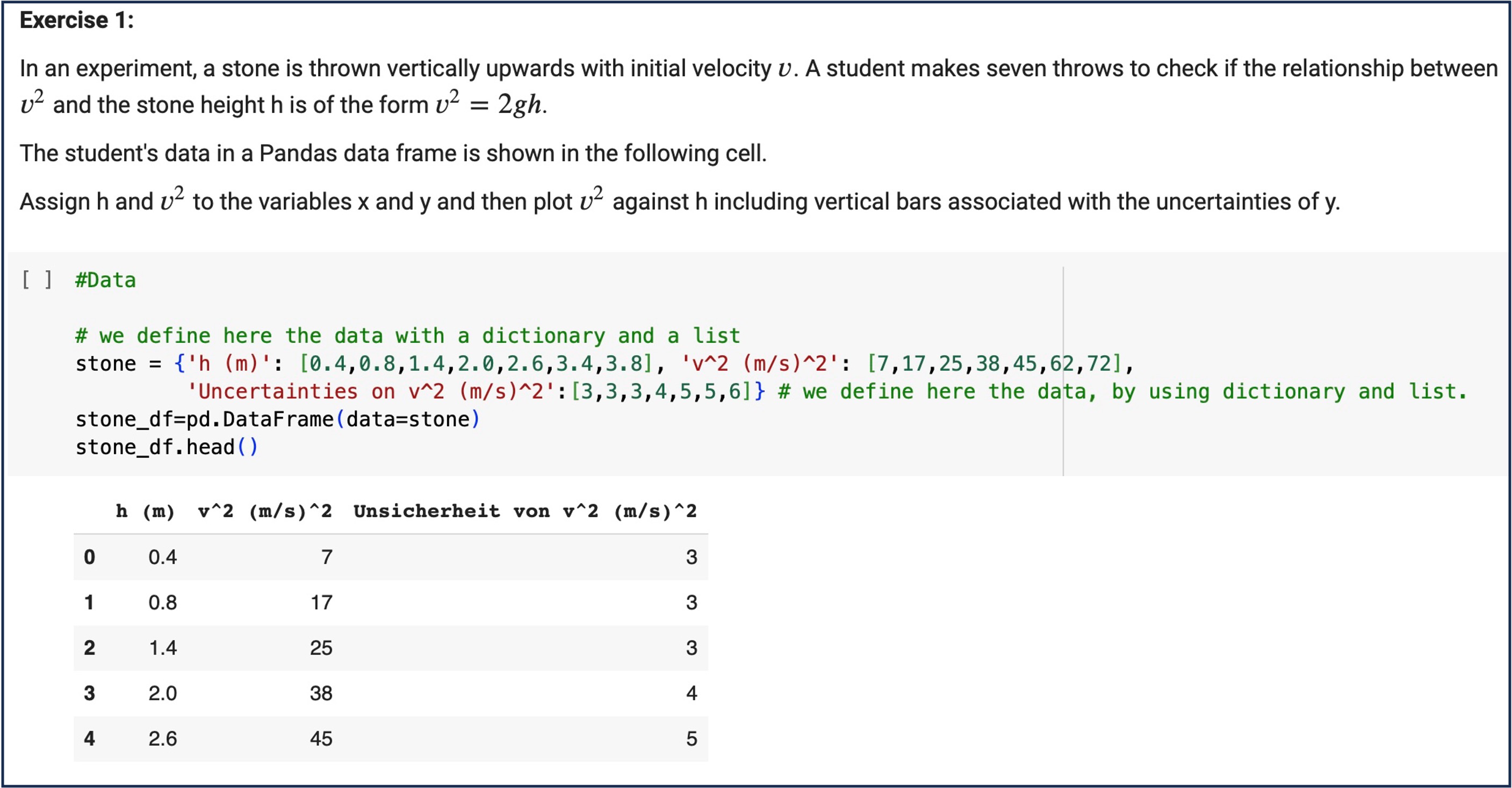}
\end{center}
\caption{\label{fig1} Snapshot of a JN with which students can practice creating a Pandas dataframe.}
\end{figure}

This exercise involves importing data into a DataFrame, performing a linearisation of the data and applying the least squares method for linear regression and finally the calculation of the reduced chi-squared. The screenshots included below (figure \ref{fig2}) show this progression. Students are first introduced to basic data handling using Pandas, progressing to more complex concepts like residuals analysis, and eventually to weighted least squares and reduced chi-square analysis. This approach not only makes the exercise scientifically significant but also allows students to see the practical application of these computational methods in a real-world physics context.

\begin{figure}
\begin{center}
\includegraphics[width=6in]{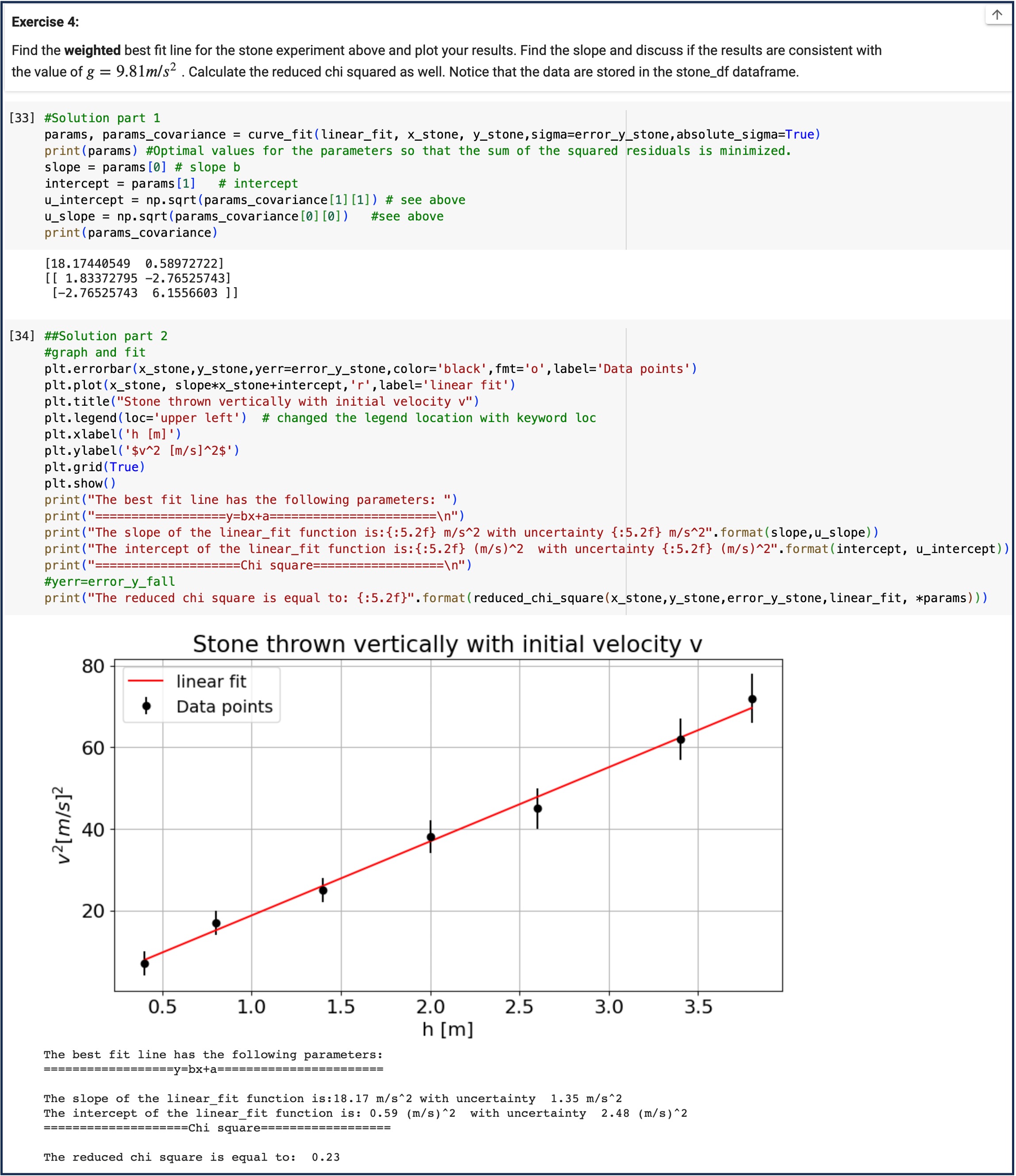}
\end{center}
\caption{\label{fig2} Snapshot of a JN with which students can practice plotting data and perform a linear regression and calculate the chi-squared.}
\end{figure}

\begin{figure}
\begin{center}
\includegraphics[width=6in]{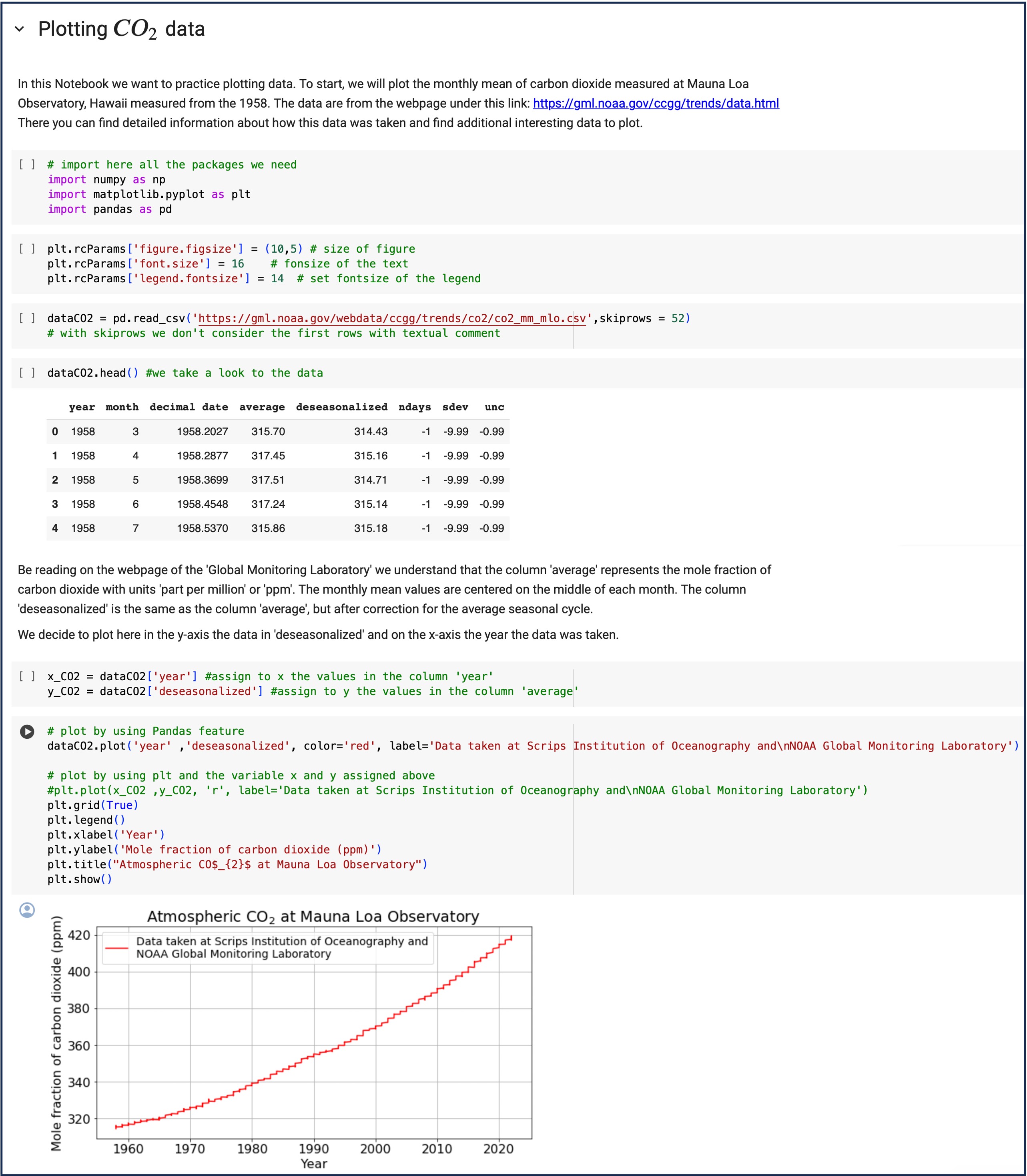}
\end{center}
\caption{\label{fig3} Snapshot of a JN with which students can practice plotting data.}
\end{figure}

\subsection{Example from JN applications series: $CO_2$ plot}
In this asynchronous application example (shown in figure \ref{fig3}), we engage students in plotting data of atmospheric $CO_2$ concentration over time, a topic of profound scientific and environmental significance.  An important initial step in the task is for students to examine the CSV data file from the scientific website NOAA Global Monitoring Laboratory. This examination is crucial as it helps students understand the file's structure and determine the starting point of the data. They learn to use the \texttt{skiprows} function in Python to correctly import the data, bypassing non-essential header information. This introduces them to a key aspect of data handling: preprocessing and cleaning data for analysis. Following this, students graphically represent the $CO_2$ data over time. This step not only illustrates the application of Python in data visualisation but also brings attention to the ongoing changes in atmospheric $CO_2$ levels.

\section{Assessment of students’ computational skills}
This section examines the assessment of students’ proficiency in using Python and Jupyter Notebooks (JNs) for data analysis in an introductory laboratory setting. The details of the assessment are described in a separate publication which is at the moment under review. We analysed students’ work using a rubric that assesses students’ skills in the two categories: 'Implement' and 'Communicate' which are aligned to our defined learning goals.  It has been applied to the various sessions illustrated  in Table 1. The rubric has 4 levels of ability, ‘missing’, ‘inadequate’, ‘needs some improvement’ and ‘near mastery’.
In figure \ref{fig:4} we present the results of the assessment, showing the percentage of students that reached each level of ability in the different in-class activities.

In the following we discuss them separately.

\subsection{First sessions: Introduction to Python (sessions S1 and S2)
}
The high scores prevalent in these sessions indicate an effective grasp of Python programming essentials early in the course, particularly in skills related to 'Implement'.

\subsection{Laboratory Experiments in the First Semester (LAB1 and LAB2)
}
Following the introductory Python sessions, students participated in two laboratory experiments, the first involving stretchable objects to explore Hooke's law \cite{smith-holmes} and the second a study on pendulum. Our analysis of students' laboratory notebooks revealed an upsurge in Python and JN utilisation, as illustrated in figure \ref{fig:4} (LAB1 and LAB2). The assessments indicate proficient use of Python in data analysis, with commendable scores in both 'Implement' and 'Communicate' categories. Notably, a progression in skill mastery is observed from LAB1 to LAB2. Python and JN, while optional, were increasingly adopted by students, evident from 83\% usage in LAB1 to complete adoption in LAB2.

\subsection{Continued Proficiency in the Second Semester (S3)}
\begin{figure}[ht]
\centering
\begin{minipage}{.48\textwidth}
  \centering
  \includegraphics[width=\linewidth]{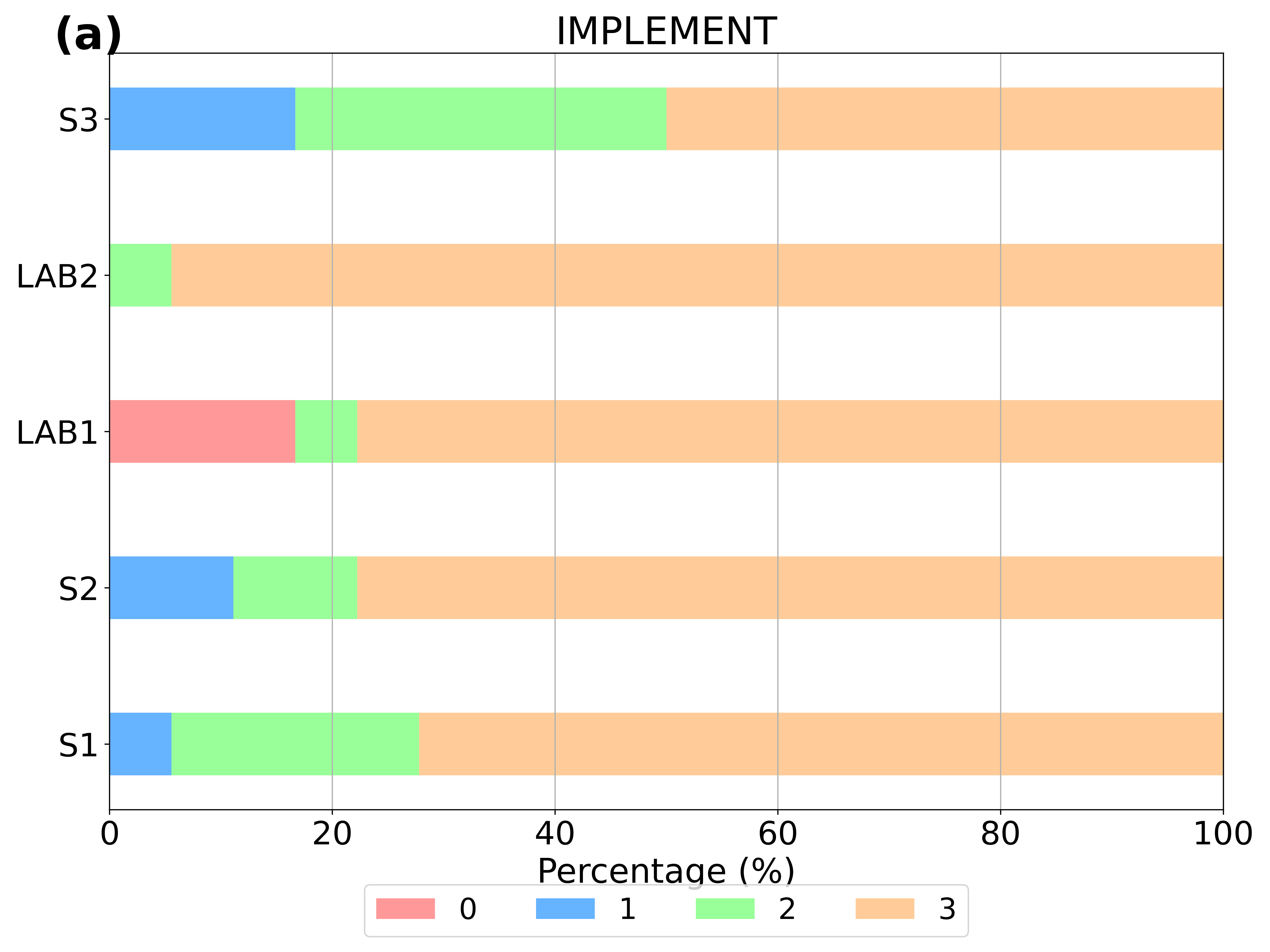}
  \label{fig:4a}
\end{minipage}%
\begin{minipage}{.48\textwidth}
  \centering
  \includegraphics[width=\linewidth]{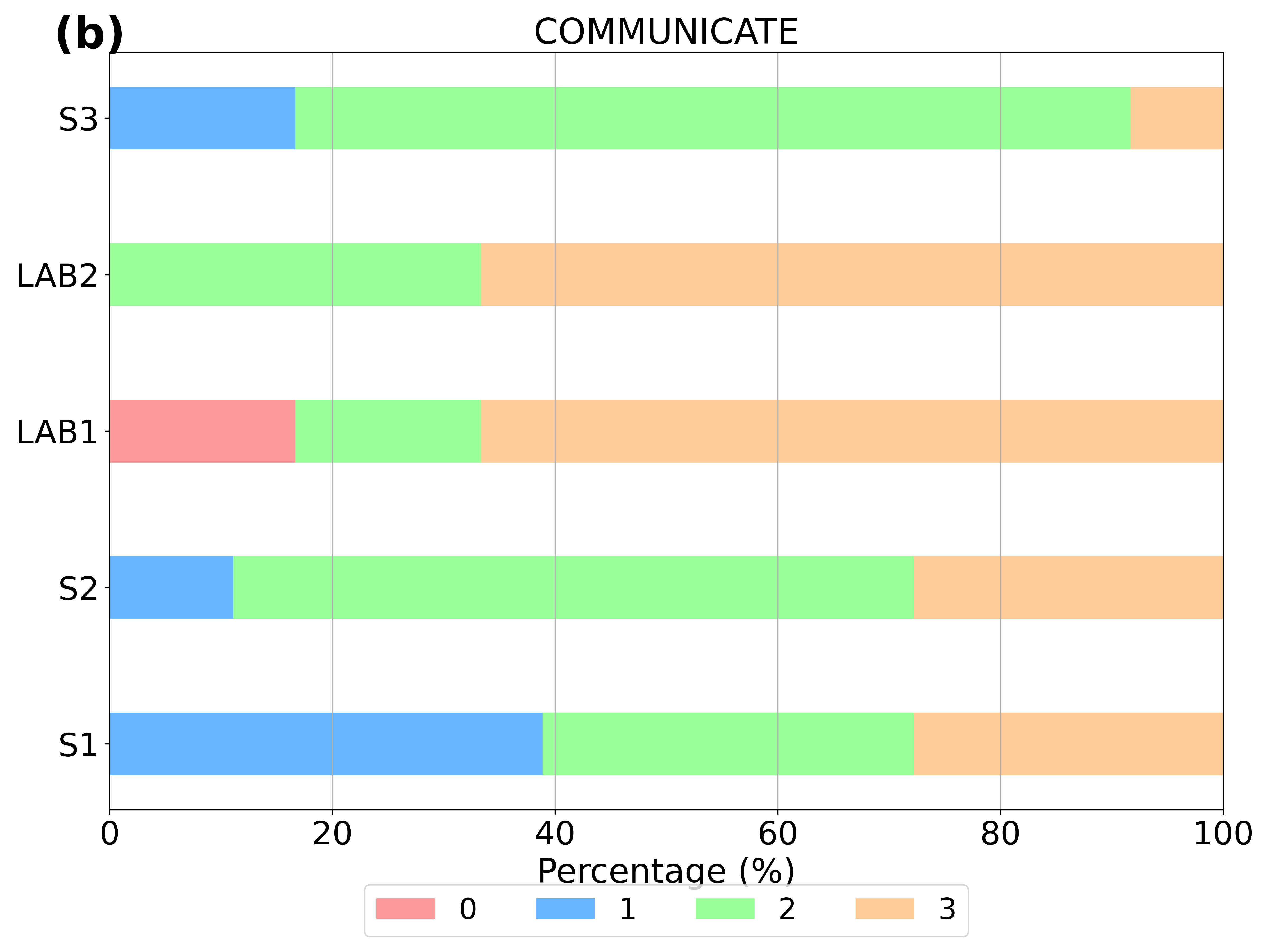}
  \label{fig:4b}
\end{minipage}
\caption{Percentage distribution of student scores in JNs across different course activities, subdivided into 'Implement' (a) and 'Communicate' (b). The Python introduction sessions (S1 and S2), the first semester laboratory experiments (LAB1 and LAB2), and the in-class exercises at the beginning of the second semester (S3) are shown distinctly. The scores range from 0 (missing), 1 (inadequate), 2 (needs some improvement), to 3 (near mastery).}
\label{fig:4}
\end{figure}

The beginning of the second semester involved revisiting Python skills through two review exercises, the outcomes of which are depicted in figure \ref{fig:4} as S3. These exercises, assigned after a three-month interruption, served as a test for the retention of computational skills. The analysis shows that about 83\% of the groups scored 2 or 3 in both 'Implement' and 'Communicate,' showing that students’ mastery of the skills imparted in the previous semester lasted over the break. However, the presence of some groups scoring 1 (inadequate) underscores the variability in skill retention.

We finally note here that in this proceedings we briefly mentioned the findings related to students' attitudes towards introducing computation in the physics laboratory course. A detailed analysis and discussion are provided in \cite{arXivPython}. Here we just refer that the analysis of the questionnaires (see table 1) suggests that students generally hold very positive views about this integration and notably maintain this positive perspective throughout the duration of our intervention. This positive outlook is crucial, as it is well-known that students' epistemological beliefs about a subject can significantly influence their learning process \cite{lising2004}. This understanding could encourage other educators to incorporate computational components into their laboratory courses.

\section{Conclusions}
In our study, we embedded computational aspects into an existing physics laboratory course, enabling students to learn programming seamlessly within the familiar course structure. The introduction of Jupyter Notebooks played a crucial role, not only in equipping students with important programming skills for their academic and professional futures but also improving their expertise in data analysis.

This methodology proved effective, as shown by the assessments that were in line with the computational learning goals we set. 
A shift from using preset data analysis tools to applying Python for tailored data analysis was well-received by students, who kept using the learned skills in the following semester. 
This positive reception is significant in the context of introducing research-based changes, which often encounter student resistance \cite{Deslauriers}.

Despite these successes, it's important to acknowledge the limitations of our study. The focus on group-based assessments may not fully reflect individual learning outcomes, and the small sample size limits the generalizability of our results. Our study's context, involving first-year physics students, some with prior programming experience, might also differ in other educational settings.

Among our future directions, we plan to continue monitoring students to assess the long-term impacts and benefits of our intervention, and to reinforce the computational skills acquired by providing further practice opportunities in subsequent laboratory courses. Additionally, we recognize the need to adapt our teaching strategies to the evolving educational landscape, particularly considering the advent of of AI programming support, assisted by Large Language Models (LLMs) like ChatGPT. 
These AI tools are rapidly transforming how we approach teaching and learning. A balanced approach is essential. It's not about wholeheartedly adopting these tools, but rather exploring their potential to augment learning. This exploration should focus on enhancing educational experiences while ensuring students acquire essential skills.

Our experience indicates that integrating Python into physics laboratory courses effectively improves computational skills while maintaining the course structure.

\end{document}